\documentclass[runningheads,citeauthoryear]{apinv43}
\usepackage{epsfig,cite,graphics}
\usepackage[utf8]{inputenc}
\usepackage{xcolor}

\begin{document}

\title{Interstellar extinction toward MWC~148 }
\titlerunning{MWC~148 }
\author{R. Zamanov\inst{1}, I. Stateva\inst{1}, J. Marti\inst{2}, K. Stoyanov\inst{1}, 
        V. Marchev\inst{1} \\  \vskip 0.5cm }
\authorrunning{Zamanov, Stateva, Marti  et al. }
\tocauthor{R. Zamanov\inst{1}, V. Stateva\inst{1}, K. Stoyanov\inst{1}, 
           V. Marchev\inst{1}, J. Marti\inst{2}} 
\vskip 0.2cm 
\institute{Institute of Astronomy and National Astronomical Observatory, Bulgarian Academy of Sciences, Tsarigradsko Shose 72, BG-1784 Sofia, Bulgaria 
	\and Departamento de F\'isica, 
	     Escuela Polit\'ecnica Superior de Ja\'en, Universidad de Ja\'en, 
             Campus Las Lagunillas,  A3, 23071, Ja\'en, Spain  \newline
	     \vskip 0.2cm 
	\email{e-mails: stateva@astro.bas.bg,  kstoyanov@astro.bas.bg \\ \vskip 0.1cm }    
	} 
\papertype{Research report, submitted on xx.xx.xxxx; accepted on xx.xx.xxxx}
	

\maketitle

\begin{abstract}
We analyse high resolution optical spectra of MWC~148 (optical counterpart of the $\gamma$-ray 
source HESS J0632+057) obtained at  Observatoire de Haute Provence and  Rozhen Observatory. 
We measure equivalent widths of 7 diffuse interstellar bands and estimate the interstellar extinction 
$E_{B-V}=0.85 \pm 0.08$.
\end{abstract}
\vskip 0.2cm 

\section{Introduction}

MWC~148  (HD 259440, BD+05 1291)  was identified as the counterpart
of the variable TeV source HESS J0632+057 
(Aharonian et al. 2007,  Aragona et al. 2010;  Matchett \& van Soelen 2025). 
HESS J0632+057 belongs to a rare subclass of binary systems that emit $\gamma$-rays above 100 GeV.   
The  optical counterpart of HESS~J0632+057 
is the Be star MWC 148, which, through $Gaia$ EDR3 parallaxes (Gaia Collaboration et al. 2021) 
was estimated to be at a distance of  $1759 \pm 90$ pc (Bailer-Jones et al.  2021).  
It is a binary system consisting of a Be star and a compact object.
The secondary can be a neutron star or a black hole. 
The orbital period is $317.3 \pm  0.7$~days, 
obtained with a refined analysis of X-ray data (Adams et al. 2021). 
There is also a modulation of the very high-energy $\gamma$-ray fluxes 
with a similar period $316.7 \pm 4.4$ days. 
Its high-energy light curve features a sharp primary peak and broader secondary peak 
(Tokayer et al. 2021).
The optical emission lines of MWC~148 are almost identical 
to those of the well-known Be star $\gamma$~Cas (Zamanov et al. 2016),
which indicates that the circumstellar discs and the orbital periods are 
comparable. 

For the interstellar extinction toward MWC~148 NASA/NED 
(Madore et al. 1992) extinction calculator 
gives an upper limit $E_{B-V} \le 0.6$ magnitude.
From other side the recent catalogue of GAIA gives a considerably larger value 
$E_{B-V} = 1.090$  (Gaia Collaboration et al. 2021). 
Here we deal with the diffuse interstellar bands (DIBs)  
visible in the high resolution optical spectra 
and estimate $E_{B-V}$ towards MWC~148.
 \begin{figure}    
   \vspace{8.3cm}     
   \includegraphics{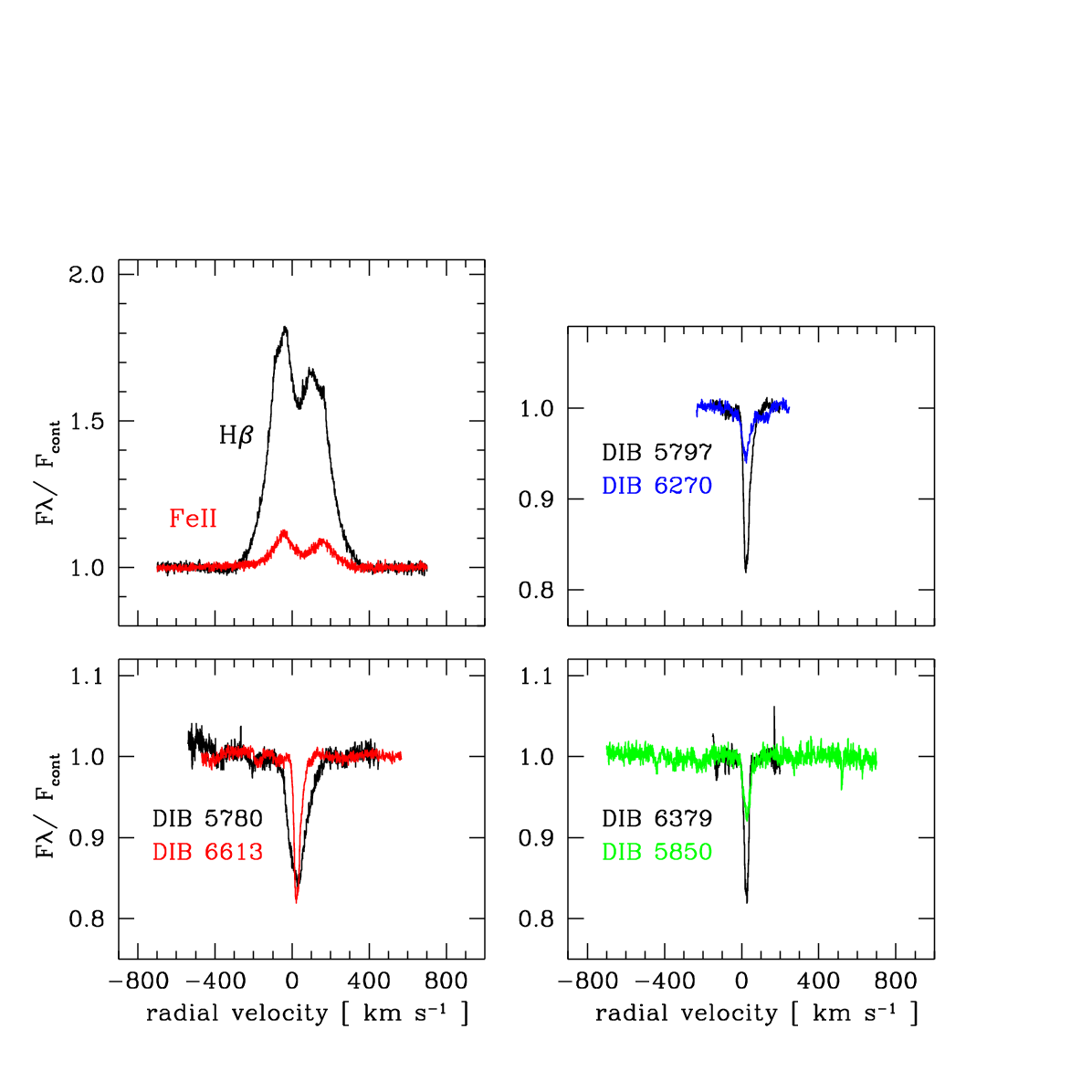}  
   \caption[]{Profiles of the $H\beta$ and FeII~5316 emission lines  
              together with
	      diffuse interstellar bands DIB 5780, 6613, 5797, 6270, 6379, 5850. }
  \label{f.1}      
  \end{figure}        
\begin{table*}
\caption{Journal of spectral  observations of MWC~148.
In the columns are given the observatory, 
date, UT of the start of the exposure, the exposure time, 
and signal-to-noise ratio around 6600~\AA.}             
\centering
\begin{tabular}{l c | c c c | c c | c c c c  ccc ccc} 
\\   
\# \hskip 0.5cm  observatory   & &  &   date-obs	 & &  &   exposure   &  & S/N & \\  
$^{}$ \hskip 0.8cm    spectrograph  & &  &   		 & &  & 	     &  &     & \\
               & &  &   		 & &  & 	     &  &     & \\  
\#1 \hskip 0.5cm OHP  SOPHIE   & &  &  2011-11-23  03:41 & &  &       60 min &  & 160 & \\
\#2 \hskip 0.5cm OHP  SOPHIE   & &  &  2011-12-08  00:26 & &  &       60 min &  & 145 & \\
\#3 \hskip 0.5cm OHP  SOPHIE   & &  &  2011-12-21  00:50 & &  &       60 min &  & 150 & \\
\#4 \hskip 0.5cm OHP  SOPHIE   & &  &  2011-12-29  00:55 & &  &       60 min &  & 110 & \\
\#5 \hskip 0.5cm OHP  SOPHIE   & &  &  2012-01-03  00:22 & &  &       60 min &  & 140 & \\
\#6 \hskip 0.5cm Rozhen ESpeRo & &  &  2022-04-14  19:04 & &  &       40 min &  &  45 & \\  
\#7 \hskip 0.5cm Rozhen ESpeRo & &  &  2024-02-27  18:23 & &  &       90 min &  &  40 & \\
\#8 \hskip 0.5cm Rozhen ESpeRo & &  &  2024-01-22  19:02 & &  &       45 min &  &  55 & \\ 
\hline                                           		    
  \end{tabular}                                                  
  \label{t.jou}
\vskip 0.5cm 
\caption{Equvalent widths  of DIBs in the spectra of MWC~148. 
The last row gives the full width of the half maximum (FWHM) for each DIB. }             
\centering
\begin{tabular}{l c c c | c c | c c c c  ccc ccc} 
\hline    
                   & 	 &	     &          &	       &	     &	    &	       & \\  
\#                 & $EW_{5780}$& $EW_{5797}$  & $EW_{6614}$  & $EW_{6270}$ &  $EW_{6379}$  &  $EW_{5850}$ & $EW_{6196}$ &\\                   &   [\AA]     &  [\AA]       &  [\AA]       &   [\AA]     &   [\AA] &  [\AA]  &  [\AA] & \\  
\#1 \hskip 0.5cm   & 0.3425   & 0.1444  & 0.1706   &  0.0630   &  0.1020   &  0.0656   &  0.0353  & \\
\#2 \hskip 0.5cm   & 0.3166   & 0.1379  & 0.1696   &  0.0555   &  0.1061   &  0.0604   &  0.0354  & \\
\#3 \hskip 0.5cm   & 0.3074   & 0.1563  & 0.1746   &  0.0501   &  0.1015   &  0.0695   &  0.0353  & \\
\#4 \hskip 0.5cm   & 0.2887   & 0.1278  & 0.1719   &  0.0647   &  0.1038   &  0.0681   &  0.0422  & \\
\#5 \hskip 0.5cm   & 0.3102   & 0.1405  & 0.1876   &  0.0756   &  0.0941   &  0.0688   &  0.0405  & \\
\#6 \hskip 0.5cm   & 0.3345   & 0.1517  & 0.174    &	       &	   &		&	   & \\  
\#7 \hskip 0.5cm   & 0.3328   & 0.1420  & 0.157    &	       &	   &		&	   & \\  
\#8 \hskip 0.5cm   & 0.3424   & 0.1491  & 0.164    &	       &	   &		&	   & \\  
                   &	      & 	&	   &	       &	   &	       &	  & \\
\hline                                              		
FWHM [km s$^{-1}$] & $107\pm2$&	$38\pm1$& $42\pm1$ &  $49\pm2$ & $26\pm1$  & $40\pm2$ &  $20\pm1$ & \\ 
\hline
  \end{tabular}                                                  
  \label{t.1}
\vskip 0.5cm
\caption{Estimated $E_{B-V}$ from EWs of DIBs in the spectra of MWC~148:    
(1) using ODR  and (2) OLS  fitting of Puspitarini et al. (2013).  }         
\centering
\begin{tabular}{l | c | c | cc | cc  ccc  ccc ccc} 
\hline
 	      &          &  $E_{B-V}$       &  $E_{B-V}$       & \\ 
              &   DIB    &  (1)             &  (2)             & \\  
	      &          &                  &                  & \\ 
	      &	DIB~5780 & $0.73 \pm 0.04$  &  $0.78 \pm 0.05$ & \\
	      & DIB~5797 & $0.91 \pm 0.07$  &  $0.81 \pm 0.06$ & \\
	      & DIB~6614 & $0.89 \pm 0.04$  &  $0.90 \pm 0.04$ & \\
	      & DIB~6270 & $0.95 \pm 0.15$  &  $0.75 \pm 0.11$ & \\
	      & DIB~6379 & $1.06 \pm 0.05$  &  $0.99 \pm 0.04$ & \\
	      & DIB~5850 & $0.83 \pm 0.05$  &  $0.76 \pm 0.04$ & \\
	      & DIB~6196 & $0.99 \pm 0.09$  &  $0.74 \pm 0.07$ & \\
 	      &          &                  &                  & \\ 
\hline                                           		    
 	      &          &                  &                   & \\ 
 	      &  average & $0.909\pm 0.108$ & $0.820 \pm 0.092$ & \\ 
 	      &          &                  &                   & \\ 
  \end{tabular}                                                  
  \label{t.2}
\end{table*}

\section{Observations} 
\label{s.obs}

In this study  we  use  five optical spectra obtained with the SOPHIE spectrograph 
(Perruchot et al. 2008) 
mounted on the 1.93m telescope of the Haute-Provence Observatory (OHP), France.
The spectra are downloaded from the SOPHIE Archive which gives access 
to the observations obtained with the spectrograph since 
it started operations in August 2006.  
Additionally,  three spectra have been secured with 
the ESpeRo echelle spectrograph (Bonev et al. 2017) mounted on the 2.0m RCC telescope 
of the Rozhen National Astronomical Observatory, Bulgaria. 
The SOPHIE spectrograph has a  resolution of $\sim 75~000$
and ESpeRo -- of $\sim 30~000$. 
Journal of observations is given in Table~\ref{t.jou}.

On the SOPHIE spectra we identified 12 DIBs -- 
5780.38~\AA,  5797.06~\AA,  6613.62~\AA,  6269.85~\AA,  6379.32~\AA,  5849.81~\AA,  6195.98~\AA,  
4726.83~\AA,  4762.61~\AA, 4963.88~\AA,  6660.71~\AA,  6699.32~\AA. 
The equivalent widths (EWs) of seven of them were measured with good accuracy.
On the Rozhen spectra we were able to measure the EWs of three DIBs. 


\section{Results}

For each DIB the local continuum was drawn using  a section of about 20~\AA. 
The EWs of the DIBs were measured using $splot$ routine of $IRAF$. 
The profiles of DIBs are shown in Fig.~\ref{f.1}. 
For illustrative  
purposes are also plotted the profiles of the H$\beta$
and FeII~5316 emission lines. 
The EW of DIBs in the spectra of MWC~148 are given in Table~\ref{t.1}.
The typical error is 5\%-10\%. 


To calculate the interstellar extinction we use the results in Puspitarini et al. (2013), 
where are given the relations between the EWs of the DIBs and the interstellar extinction $E_{B-V}$. 
We will use them in the form: 
{\small
\begin{equation} 
E_{B-V}= 0.0006 + 2.5 \; EW_{5780}
\end{equation}
\begin{equation}  
E_{B-V}= 0.0086 + 2.3 \; EW_{5780} 
\end{equation} 
\begin{equation} 
E_{B-V}= 0.0291 + 5.5 \; EW_{5797}  
\end{equation}
\begin{equation}  
E_{B-V}= 0.0203 + 6.3 \; EW_{5797} 
\end{equation} 
\begin{equation} 
E_{B-V}= 0.0051 + 5.1 \; EW_{6614}  
\end{equation}
\begin{equation}  
E_{B-V}= 0.0008 + 5.1 \; EW_{6614} 
\end{equation} 
\begin{equation} 
E_{B-V}= 0.0051 + 5.1 \; EW_{6270}  
\end{equation}
\begin{equation}  
E_{B-V}= 0.0008 + 5.1 \; EW_{6270} 
\end{equation} 
\begin{equation} 
E_{B-V}= 0.0383 +  9.4 \; EW_{6379} 
\end{equation}
\begin{equation}  
E_{B-V}= 0.0359 + 10.1 \; EW_{6379} 
\end{equation} 
\begin{equation} 
E_{B-V}= -0.0073 + 11.6 \; EW_{5850},
\end{equation}
\begin{equation}    
E_{B-V}= -0.0163 + 12.7 \; EW_{5850}, 
\end{equation} 
\begin{equation} 
E_{B-V}= -0.0277 + 20.4 \; EW_{6196}
\end{equation}
\begin{equation}    
E_{B-V}= -0.0349 + 21.7 \; EW_{6196}   
\end{equation} }
In the above equations all  EWs are in \AA. 
For each band there are two equations. They are taken from
Table~2 of Puspitarini et al. (2013), where the relations are 
calculated with ordinary least square (OLS) and 
orthogonal distance relation (ODR). We use their coefficients marked with "*", 
where peculiar objects are excluded. 
The calculated value of $E_{B-V}$ for each DIB are given in Table~\ref{t.2}. 
As is visible  the ODR coefficients give of about 10\% larger value than 
OLS coefficients.  
Bearing in mind the errors of the two fits, our estimation 
for the extinction toward MWC~148 is  $E_{B-V} = 0.850 \pm 0.08$.

\section{Discussion}
\label{D.1} 

%
%

When observing astronomical objects, we have to deal with the extinction,  
i.e. the absorption and scattering of the radiation
by dust and gas between the source and the observer. 
The interstellar extinction depends on the location of the object 
and the amount of interstellar clouds between the object and the Earth.

For the interstellar extinction toward MWC~148 NASA/NED extinction calculator
gives $E_{B-V} \le 0.6$. This value is on the basis of 
the Galactic dust reddening maps provided by Schlafly \& Finkbeiner (2011).
Green et al. (2019)  and  Lallement et al. (2019)  give a similar result $E_{B-V} \le 0.5$ .
These catalogues are 3D maps of the interstellar dust reddening 
and are based on the Pan-STARRS~1 and 2MASS colours ranging wavelengths from 400 to 2400~nm. 
From other side a few recent catalogues give a considerably larger value 
$E_{B-V} = 0.854 \pm 0.066$ (Paunzen et al. 2024),  
$E_{B-V} = 0.865$ (Chen et al. 2019), 
$E_{B-V} = 1.090$ (Gaia Collaboration et al. 2021).  
The catalogue of Paunzen et al. (2024) gives
reddening estimations based on the classical photometric indices 
in the Geneva, Johnson, and Str\"omgren-Crawford photometric systems. 
Chen et al. (2019) estimate the reddening using a 3D interstellar dust
reddening map of the Galactic plane based on Gaia~DR2, 
2MASS and WISE photometry that covers the wavelength range from 400 to 2400~nm.
The Gaia~DR3 catalogue presented in Gaia Collaboration et al. (2021)
is using low-resolution spectra that cover a wide range of extinction values. 

%
%

Our estimation for the extinction toward MWC~148 ($E_{B-V} = 0.850 \pm 0.08$)
based on the EWs of seven  DIBs is in agreement 
with the values of Paunzen et al. (2024) and Chen et al. (2019). 
Our value does not correspond to some values 
obtained using the  interstellar dust reddening.
Possible reasons for the discrepancy could be that  
(1) the interstellar medium is peculiar in the direction toward MWC~148,
and/or 
(2) part of the DIBs visible in the spectra of MWC~148 
are not formed in the interstellar dust but in the interstellar gas 
or in the circumstellar environment 
(e.g. Be circumstellar disc, circumbinary material, etc.).   


\vskip 0.3cm 

{\bf Conclusions: }
Using  high resolution optical spectra 
obtained at  Observatoire de Haute Provence and Rozhen Observatory,  
we measure equivalent widths of seven diffuse interstellar bands 
and estimate the interstellar extinction  
$E_{B-V}=0.85 \pm 0.08$ toward MWC~148 (HESS J0632+057).

\vskip 0.3cm 

{\small {\bf Acknowledgments: }
This work is partly supported by grant 
"Studying the Outflows of High-Energy Sources: an observational multi-wavelength approach"
(PID2022-136828NB-C42 funded by the Spanish MCIN/AEI/ 10.13039/501100011033) 
and by {\bf National Roadmap for Scientific Infrastructure coordinated by Ministry of Education and Science of Bulgaria}.


\begin{thebibliography}{}

\bibitem{} Adams, C.~B., Benbow, W., Brill, A., et al.\ 2021, ApJ, 923, 241


\bibitem{} Aharonian, F.~A., Akhperjanian, A.~G., Bazer-Bachi, A.~R., et al.\ 2007, A\&A, 469, L1
	     
\bibitem{} Aragona, C., McSwain, M.~V., \& De Becker, M.\ 2010, ApJ, 724, 306 
	     

\bibitem{} Bailer-Jones, C. A. L., Rybizki, J., Fouesneau, M., 
             Demleitner, M., \& Andrae, R. 2021, AJ, 161, 147

\bibitem{} Bonev, T., Markov, H., Tomov, T., et al. 2017, 
             Bulgarian Astronomical Journal, 26, 67  (arXiv:1612.07226)
	     
\bibitem{} Chen, B.-Q., Huang, Y., Yuan, H.-B., et al.\ 2019, MNRAS, 483, 4277 
	     
\bibitem{} Gaia Collaboration,  Brown, A. G. A.,  Vallenari, A.,  
             Prusti, T.,  2021, A\&A, 649, A1
	     
\bibitem{} Green, G.~M., Schlafly, E., Zucker, C., Speagle, J. S., 
             Finkbeiner, D.\ 2019, ApJ, 887, 93

\bibitem{} Lallement, R., Babusiaux, C., Vergely, J.~L., et al.\ 2019, A\&A, 625, A135 

\bibitem{} Paunzen E., Netopil M., Prisegen M. et al.\ 2024, A\&A, 689, A270

\bibitem{} Perruchot, S., Kohler, D., Bouchy, F., et al.\ 2008, 
             SPIE Conference  Series, 7014, 70140J   

\bibitem{} Puspitarini, L., Lallement, R., \& Chen, H.-C.\ 2013, A\&A, 555, A25 

\bibitem{} Schlafly, E.~F. \& Finkbeiner, D.~P.\ 2011, \apj, 737, 103 

\bibitem{} Tokayer, Y.~M., An, H., Halpern, J.~P., et al.\ 2021, ApJ, 923, 17  
	      
\bibitem{} Matchett, N. \& van Soelen, B.\ 2025, MNRAS, 536, 166 

\bibitem{} Madore, B.~F., Helou, G., Corwin, H.~G., et al.\ 1992,  ASP Conf.,  25, 47 


\bibitem{} Zamanov, R., Stoyanov, K., \& Mart{\'{\i}}, J.\ 2016, 
             Bulgarian Astronomical Journal, 24, 40 (arXiv:1509.04191)


\end{thebibliography}
\end{document}